      [1999/12/01 v1.4c Il Nuovo Cimento]
\documentclass{cimento}

\usepackage{graphicx}  % got figures? uncomment this

\def\Liso{\mbox{L$_{\rm iso}$}}

\def\flux{\mbox{photons cm$^{-2}$ s$^{-1}$}}

\def\fldl{\mbox{$f$(L$_{\rm iso}$)dL$_{\rm iso}$}}

\def\ergs{\mbox{erg \ s$^{-1}$}}

\title{The Luminosity Function and Formation Rate History of GRBs}
\author{C.~Firmani\from{ins:x}\from{ins:y}\ETC,
V.~Avila-Reese\from{ins:y},
G.~Ghisellini\from{ins:x},
        \atque
A.V.~Tutukov\from{ins:z}\\
}
\instlist{\inst{ins:x} Osservatorio Astronomico di Brera, via E.Bianchi 46, I-23807 
            Merate, Italy
  \inst{ins:y} Instituto de Astronom\'{\i}a, U.N.A.M., A.P. 70-264, 04510, 
M\'exico, D.F., M\'exico
  \inst{ins:z} Institute of Astronomy, 48 Pyatnitskaya, 109017 
            Moscow, Russia}
\PACSes{\PACSit{98.70.Rz}{gamma ray bursts}}

\begin{document}

\maketitle

\begin{abstract}
The isotropic luminosity function (LF) and formation rate history (FRH) of 
long GRBs is by the first time constrained by using {\it jointly} both the 
observed GRB peak-flux and redshift distributions. Our results support 
an evolving LF and a FRH that keeps increasing after $z=2$. We discuss 
some interesting implications related to these results.
\end{abstract}

\section{Introduction}

The primordial and most extensive information on 
GRBs obtained directly from observations is the differential peak-flux, $P$, 
distribution, the so commonly called $logN-logP$ diagram (NPD). The NPD is the 
convolution of basically three factors: the intrinsic LF, the FRH and the 
cosmic volume. In the past, several attempts were done to constrain the 
LF from fits to the NPD, assuming arbitrary GRB FRHs and cosmologies. 
The results were rather poor because the complicated mixing between the 
model LF and FRH introduces a high degeneracy among these factors in the NPD.
The most direct way to infer both the LF and the FRH is based on the 
observational luminosity-redshift diagram ~\cite{ref:Schaefer,ref:Lloyd02,ref:Yonetoku}. 
Unfortunately, this strategy is so far limited by the small number of GRBs with 
known or inferred $z$, by the bias introduced by the given method to infer
$z$, and by the sensitivity limit related to the $z$ estimate. We 
propose~\cite{ref:FAGT} a new strategy for constraining the GRB LF and FRH 
based on the joint use of the NPD and the observed or inferred GRB differential 
$z$ distribution (the $N-z$ diagram, NZD).

\section{The Method and the Data}

The differential NPD and NZD are modelled by seeding at each $z$ a large number 
of GRBs with a given rate, $\dot{\rho}_{\rm GRB}$, and LF, \fldl, and then by 
propagating the flux of each source to $z=0$. We use the popular $\Lambda$CDM 
cosmology with $\Omega_m$=0.29, $\Omega_{\Lambda}=h=0.71$. Here \Liso\ in 
the rest frame is defined as \Liso=$\int_{\rm 30keV}^{\rm 10000keV}ES(E)dE$,
where $S(E)$ is the Band~\cite{ref:Band} energy spectrum with the (average) 
parameters taken from ~\cite{ref:Preece}. The break energy at rest, $E_b$, 
is assumed either 
constant (512 keV ~\cite{ref:PM01}) or dependent on \Liso~\cite{ref:Yonetoku} 
($E_b = 15\ (\Liso/10^{50}\ \ergs)^{0.5}$ keV). The sensitivity band at 
$z=0$ is fixed to the 50--300 keV range of BATSE, . 

We explore two models for the LF, the single and double power laws (SPL
and DPL, respectively), and two cases, one with a non evolving LF, 
and another one where \Liso\ scales with $z$ as $(1+z)^{\delta}$
~\cite{ref:Lloyd02,ref:Yonetoku}. For $\dot{\rho}_{\rm GRB}$, we adopt 
the bi-parametric star formation rate (SFR) function, $\dot{\rho}_{\rm SF}(z;a,b)$, 
given in ~\cite{ref:PM01} multiplied by a normalisation factor $K$, and by a 
function $\eta (z;c)$ that allows to control, through its parameter $c$, the 
growth or decline of $\dot{\rho}_{\rm GRB}$ at $z>2$.  The strategy is
to constrain seven parameters (3 of LF, 3 of FRH and $\delta$, for details 
see~\cite{ref:FAGT}) 
by applying a {\it joint} fit of the model predictions to the 
observed NPD and NZD (including their errors). The fitting is based on 
an extension of the Levenberg-Marquardt method to find the least total
$\chi^2 = \chi_{\rm NP}^2 + \chi_{\rm NZ}^2$.   

We use the widest sample to date for the NPD~\cite{ref:Stern}. It consists 
of a collection of (i) 3255 BATSE GRBs longer than 1 s, and with a $P$ limit 
appropriately extended down to 0.1 \flux, and (ii) a sample of bright 
{\it Ulysses} GRBs with $P$ up to $\sim 300$ \flux. For the NZD, we use
two samples. One consists of a set of 220 BATSE GRBs with $z'$s inferred 
by using the luminosity-variability empirical relation ~\cite{ref:FR-R00},
and the other comprises 33 GRBs with known $z$. The latter is corrected
for several selection effects. 

\begin{figure}
\vspace{8.1cm}
\includegraphics{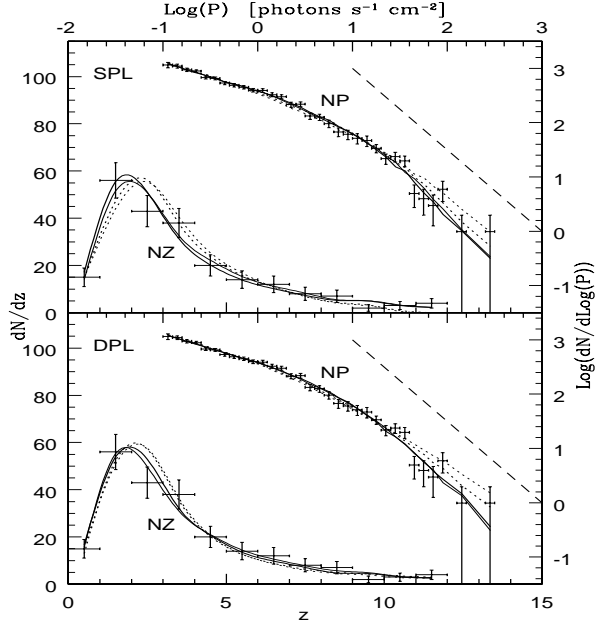}
\caption{{\it Top panel:} Peak flux differential distribution (NP) with 
the axis in the top-right part, and $z$ differential distribution 
(NZ) with the axis in the bottom-left part, both for a SPL LF. Error 
bars show the NP data from ~\cite{ref:Stern} and the NZ data according to 
~\cite{ref:FR-R00}, respectively. Dotted lines are for models without 
evolution, while solid lines are for models with the evolution 
parameter $\delta$ optimised. Thin and thick lines identify the cases 
with E$\rm_b$=511 keV or E$\rm_b$ depending on \Liso, 
respectively. Dashed straight line is the -3/2 uniform distribution 
(Euclidean) behaviour in the NPD. 
{\it Bottom panel:} Same as in top panel but for a DPL LF.}
\end{figure}

\section{Results and Discussion}

We have run several combinations of models, with an SPL or DPL LF,
adopting an $E_b$ constant or dependent on \Liso, and either without
evolution ($\delta=0$) or including  $\delta$ among the parameters
to optimise. The obtained LFs and FRHs are shown in Fig. 1 (curves)
together to the observational data (error bars).  A detailed analysis 
of the fittings of these 
different models to the NPD and NZD data is presented in ~\cite{ref:FAGT}.
In the following, we highlight some of the results from this analysis:

{\it GRB LF and jet angle distribution.} As seen in Fig. 1 models with 
non-evolving LFs (SPL or DPL, dotted curves) give in general poor fits to 
the NPD and NZD. The best joint fits are obtained for models with evolving LFs
($\propto (1+z)^{\delta}$, solid lines). The optimal values we find for 
$\delta$ are $1.0 \pm 0.2$.
Note that the increasing of \Liso\ with $z$ increases the probability to observe GRBs 
from very high $z'$s. The fits are slightly better for the (evolving) SPL LFs
than for the DPL ones. The best range of slopes of the SPL LF is 
$\gamma=1.57\pm 0.03$ ($\fldl\propto \Liso^{-\gamma}$). If the LF is related
to collimation effects, we have compared our results with the 
{\it universal structured jet} and
with the {\it quasi--universal Gaussian structured jet} models
~\cite{ref:FAGT}.
Our results imply an intermediate case between the 
universal  structured jet model with $\epsilon(\theta)\propto \theta^{-2}$ and 
the quasi--universal Gaussian structured jet model. For the uniform jet model, 
the jet angle 
distribution covers an indicative range between 2$^\circ$ and 15$^\circ$ at $z=1$.
These results correspond to the case of a luminosity-depending $E_b$ in the model
LFs, but we find that for $E_b=$const the differences are not significant.     

{\it GRB FRH and the connection with the cosmic SFR.} The models that best
fit the observed NPD and NZD imply not only an evolving LF but also FRHs, 
$\dot{\rho}_{\rm GRB}(z)$, which steeply increase (by a factor of $\sim 30$) 
from $z=0$ to $z\approx 2$ and then continue increasing gently up to $z\sim 10$
as $(1+z)^{1.4}$ and $(1+z)$ for the SPL and DPL LFs, respectively. This 
behaviour of the GRB FRH is qualitatively similar to the one of the cosmic SFRH.
Fig. 2 shows a comparison of the cosmic SFRH traced by rest-frame
UV luminosity (corrected by dust obscuration) with the GRB FRH translated
to SFRH under the assumption of a non evolving Initial Mass Function 
(IMF). The shaded area corresponds to the range of GRB models (including the 
uncertainties) with evolving ($\delta=1$) SPL and DPL LFs that best fit 
the NPD and NZD data.

\begin{figure}
\vspace{6.5cm}
\includegraphics{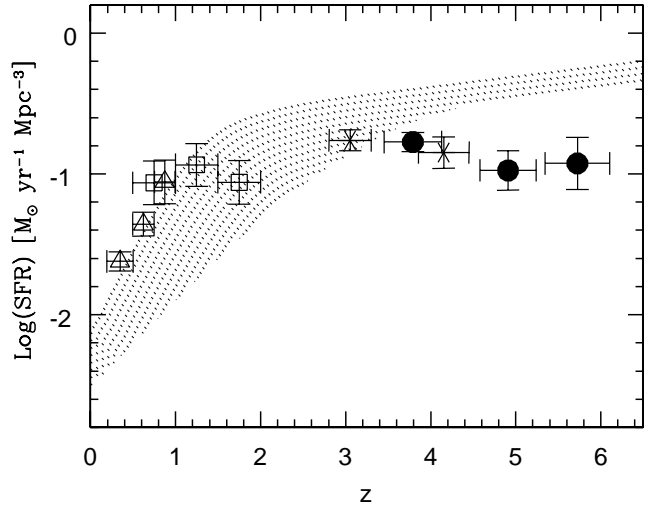}
\caption{Comparison between the observed cosmic SFRH traced by the rest-frame 
UV luminosity from \cite{ref:Giav} (dots with error bars) and from \cite{ref:Ellis} 
(triangle) and the SFRH obtained from GRB FRH properly 
normalised (shaded area, see text)}
\end{figure}

From $z=2-3$ to $z=6$, the cosmic SFRH traced by UV luminosity decreases 
slightly~\cite{ref:Giav} or strongly~\cite{ref:Ellis}, while the SFRH linked 
to GRBs keeps increasing. This difference, if proved in future studies, could 
be related to two effects. (i) A significant contribution to the cosmic SFRH 
at $z>2$ seems to come from sources emitting strongly in the rest-frame far 
infrared/submillimetre (not seen in UV).  GRBs could be also tracers of the 
dust-enshrouded SFRH of these galaxies (e.g., ~\cite{ref:R-R02}). (ii) The 
IMF in low-metallicity gas (high $z'$s) could be biased toward higher
masses compared to the present-day IMF, favouring in this way an increasing
GRB formation rate with z. 
From $z=1-2$ to $z=0$ the UV-luminosity SFRH decreases by less than an order 
of magnitude, while the SFRH linked to GRBs decreases roughly by
a factor of $30$.
Due to the uncertainty on the z determination any definitive conclusion 
about this difference has to be supported by a more extended sample 
with know z for NZD. 

{\it Implications for the progenitors.} Our best models give a {\it true} 
(after collimation effect correction) GRB FR of $\sim$ 5 10$^{-5}$ yr$^{-1}$ for 
the Milky Way. Based on astronomical arguments we argue that such a FR is 
close to that of close binary systems consisting of a WR star and a possible
massive BH, with periods of hours. These systems are able to 
generate a massive Kerr BH after the 
SNIb/c explosion of the WR (helium) star. The observational counterparts of 
these potential GRB progenitors should be luminous X-ray binaries 
(e.g., Cyg X-3), which are estimated to be only a few at present 
in the MW.

{\it Implications for cosmology.} The finding of the LF properties will shed
new light on the connection between the {\it true} (collimation corrected) 
energetics of GRBs and their spectral features. This connection, studied for the
first time by Ghirlanda et al. ~\cite{ref:GGL,ref:GGLFb}, 
makes GRBs powerful {\it standard candles} which hopefully will allow
to explore in few years the geometry and the kinematics of the universe beyond
$z=10$ ~\cite{ref:GGLF,ref:FGGA,ref:GGF}.

\end{document}